\documentclass[reprint, superscriptaddress, floatfix, amsmath,amssymb, aps, prx, noeprint]{revtex4-1}

\usepackage[utf8]{inputenc}
\usepackage{amsmath,esint}
\usepackage{mathtools}
\usepackage[colorlinks=true,linkcolor=blue,citecolor=blue,urlcolor=blue]{hyperref}
\usepackage{amsfonts}
\usepackage{amssymb}

\usepackage{textcomp}
\usepackage{empheq}
\usepackage{siunitx}
\usepackage[titletoc,title]{appendix}
\usepackage{graphicx}
\usepackage{dcolumn}
\usepackage{subcaption}
\usepackage{xcolor,soul}

\renewcommand{\Re}{\operatorname{Re}}
\renewcommand{\Im}{\operatorname{Im}}

\newcommand{\citeasnoun}[1]{Ref.~\cite{#1}}

\newcommand{\Figref}[1]{Figure~\ref{fig:#1}}
\newcommand{\figref}[1]{Fig.~\ref{fig:#1}}
\renewcommand{\eqref}[1]{Eq.~(\ref{eq:#1})}
\newcommand{\Eqref}[1]{Equation~(\ref{eq:#1})}

\newcommand{\eqrefrange}[2]{Eqs.~(\ref{eq:#1})--(\ref{eq:#2})}
\newcommand{\vect}[1]{\boldsymbol{\mathbf{#1}}}

\newcommand{\secref}[1]{Sec.~\ref{sec:#1}}

\newcommand*{\xv}{\mathbf{x}}

\newcommand*{\epso}{\varepsilon_0}

\newcommand*{\ts}{\textsubscript}

\begin{document}

\title{Optical Materials for Maximal Nanophotonic Response}

\author{Hyungki Shim}
\affiliation{Department of Applied Physics, Yale University, New Haven, Connecticut 06511, USA}
\affiliation{Energy Sciences Institute, Yale University, New Haven, Connecticut 06511, USA}
\affiliation{Department of Physics, Yale University, New Haven, Connecticut 06511, USA}
\author{Zeyu Kuang}
\affiliation{Department of Applied Physics, Yale University, New Haven, Connecticut 06511, USA}
\affiliation{Energy Sciences Institute, Yale University, New Haven, Connecticut 06511, USA}
\author{Owen D. Miller}
\affiliation{Department of Applied Physics, Yale University, New Haven, Connecticut 06511, USA}
\affiliation{Energy Sciences Institute, Yale University, New Haven, Connecticut 06511, USA}

\begin{abstract}
    This article reviews the material properties that enable maximum optical response. We highlight theoretical results that enable shape-independent quantification of material ``figures of merit,'' ranging from classical sum rules to more recent single-frequency scattering bounds.  A key delineation at optical frequencies is between polaritonic materials that support highly subwavelength resonances and dielectric materials that can have vanishingly small loss rates. We discuss the key metrics that enable comparisons both within these material classes and between them. We discuss analogous metrics for 2D materials, and point to applications for which rigorous comparison can be made between bulk- and 2D-material approaches. The various results highlight the synergy between materials discovery and theoretical nanophotonic bounds, and point to opportunities in achieving new extremes in light--matter interactions.
\end{abstract}

\maketitle

\section{Introduction}
As material discovery proceeds at a rapid pace~\cite{Goldberger2003a,Yaghi2003a,VanHest2005a,Macak2005a,Novoselov2005a,Geim2007a,Ayari2007a,Pacil2008a,Hsieh2008a,Dean2010a,Mak2010a,Moore2010a,Osada2012a,Wang2012a,Geim2013a,Liu2014b,Borisenko2014a,Lv2015a,Saidaminov2015a,Bhimanapati2015a,Tan2017a}, and atomistic control promises the possibility of ``designer'' materials~\cite{Frenkel2002,Barredo2016b,Kim2016b,Barredo2018b}, there is a fundamental question to be answered: what material properties should the optical-materials community aim to synthesize? In the field of nanophotonics, for many applications the goal is to maximize the interaction of light with matter, manifested by absorption and/or scattering~\cite{Gobin2007b,Huang2009c,Anker2009b,Chang2010b,Pu2010b,Hsu2014b}, quality factor~\cite{Noda2000c,Michler2000c,Loncar2003b,Reithmaier2004b,Tanabe2005b,Hennessy2007b,Srinivasan2007b}, spontaneous-emission rate~\cite{Nie1997c,Bardhan2009b,Callahan2012,Eggleston2015c}, and related response functions~\cite{Laroche2006b,Martin-Cano2010b,Gonzalez-Tudela2011b,Sarkar2012b}, for frequency bands ranging from the near ultraviolet to the far infrared. In this Review, we survey the key metrics that have been identified for maximum optical response. Across the broad landscape of polaritonic, dielectric, and 2D materials, we use experimental optical-constant data in tandem with these metrics to identify especially promising materials and material characteristics at optical frequencies.

At any frequency, one can divide the landscape of nonmagnetic materials into two categories: those whose permittivities have negative real parts, thereby supporting quasistatic plasmonic and polaritonic resonances (with caveats discussed below), and those with positive real parts. Following standard terminology, we call the former ``polaritonic'' materials and the latter ``dielectric'' materials, though by this definition the category a physical material belongs to often changes with the frequency of interest. The reason for this delineation is the significant differences in the resonator properties of the two materials. For polaritonic materials (\secref{polaritonic}), it is possible and typically desirable to support quasistatic resonances with extremely subwavelength confinement of electromagnetic waves~\cite{Maier2007,Ozbay2012,Caldwell2015,Khurgin2015}, with a length scale decoupled from the free-space wavelength and a resonant frequency determined by the material permittivity and depolarization factor associated with the shape. The dominant loss mechanism is dissipation (absorption), and for materials with electric susceptibilities $\chi(\omega)$, there are two key metrics pertaining to dissipation: an ``inverse resistivity'' $|\chi|^2/\Im \chi$ (\citeasnoun{Miller2016}), and a material quality factor proportional to $\left[\partial (\Re \chi)/\partial \omega\right] / \Im \chi$ (\citeasnoun{Wang2006}). By contrast, dielectric materials (\secref{dielectric}) require patterning at sizes at the scale of wavelength~\cite{Krasnok2012,Fu2013,Kuznetsov2016}, and radiative coupling (e.g. surface roughness) is typically the dominant loss mechanism. For these materials, a sum rule (generalized from Ref.~\cite{Barnett1996}) dictates that the all-frequency response is constrained by the real part of the refractive index, $n$, which becomes a key metric of interest. In \secref{comp}, we review two known ways to compare \emph{between} the two material categories: a sum rule that equates total scattering response to the total number of electrons in the scatterer~\cite{Gordon1963,Yang2015}, and recent ``power-bandwidth'' limits that enable comparison over any bandwidth from 0 to $\infty$~\cite{Shim2019}. We include comparisons between the two and show that it \emph{is} possible in many scenarios to quantitatively determine whether polaritonic or dielectric approaches are optimal. Finally, looking forward, we examine new avenues of exploration at the intersection of optical-material synthesis and maximum nanophotonic response (\secref{forward}).

Given the excitement over recent material breakthroughs, there are a few recent reviews surveying various aspects of the field~\cite{Naik2013,Caldwell2015,Zhong2015,Kuznetsov2016,Basov2016a,Bhimanapati2015a,Tan2017a,Low2017,Foteinopoulou2019}. In this review, we include and emphasize only those material characteristics that can be shown to be globally optimal for some application; moreover, we require such optimality to be \emph{independent} of the underlying structure or geometry of the system. We do not include analyses or properties that are only true for, e.g., spheres or planar surfaces; instead, we show that in fact there is now extensive theoretical understanding about structure-independent optimal material properties.


\section{Polaritonic Materials} 
\label{sec:polaritonic}
Polaritonic materials at optical frequencies benefit from the strong coupling of light to free electrons, but at the cost of significant absorption. To what extent can absorption losses be avoided, for large scattering or high-$Q$ resonances? In this section we highlight two results that answer this question: bounds on the largest single-frequency response possible (\secref{MSFR}) and bounds on the highest quality factor possible (\secref{QF}), yielding two metrics ($|\chi|^2 / \Im\chi$ and $\left[\partial (\Re \chi)/\partial \omega\right] / \Im \chi$, respectively, for bulk materials with susceptibilities $\chi$) by which all polaritonic materials can be compared. We also highlight the important role the \emph{real} part of the permittivity plays to determine the feasibility of achieving high-confinement polaritonic resonances (\secref{RPP}). In \secref{radeff} we highlight general bounds for maximum response in a regime of high-radiative-efficiency plasmonics, and hybrid dielectric--metal structures that offer a combination of high efficiency and large response, approaching their respective bounds. Finally, we discuss the important role \emph{nonlocality} plays at small length scales, and known bounds that incorporate the relevant nonlocal parameters (\secref{NL}).

\subsection{Maximal Single-frequency Response}
\label{sec:MSFR}
It has long been recognized that reducing material ``loss'' is critical for many plasmonics applications, yet there are many possible loss rates to choose from: the imaginary part of the susceptibility, the imaginary part of the refractive index, the real part of a material's conductivity, the inverse of its real resistivity, etc. One way to frame the question is in the context of scattering problems: given an external excitation, what is the largest possible response (absorption, scattering, etc.) from a given material? \citeasnoun{Miller2016} develops a systematic answer to this question, independent of particulars of the geometrical patterning. The idea stems from consideration of the polarization currents $\vect{P}$ excited within the material. The rate at which energy is absorbed is proportional to $\int_V \left(\Im \chi\right) |\vect{E}|^2 = \left(\Im \chi / |\chi|^2\right) \int_V \left|\vect{P}\right|^2$, and is proportional to the square of the magnitude of the polarization currents. By contrast, the well-known optical theorem~\cite{Newton1976,Lytle2005,Jackson2007,Bohren2008} dictates that the total extinction of the incident field, via absorption or scattering, must be proportional to the imaginary part of a scattering amplitude, which increases only linearly with the induced polarization currents. Extinction must be larger than absorption, and yet absorption increases more rapidly with the magnitude of the polarization currents. Constraining absorption to be smaller than extinction, which is equivalent to requiring that scattered power be non-negative, thereby imposes a limit on the largest polarization currents that can be excited in any particular material. This constraint can then be used to identify bounds on many optical-response functions. The exact form of the bound depends on the application, but within each bound there is a ``material figure of merit'' $f(\omega)$ that dictates increased possible response based on the material properties. (A similar analysis~\cite{Miller2017a} can be done for the induced surface currents in 2D materials with local surface conductivities $\sigma(\omega)$.) For any polaritonic material, the material metric is given by
\begin{equation}
		f(\omega) =
    \begin{dcases}
        \frac{|\chi(\omega)|^2}{\Im{\chi(\omega)}} & \text{3D / bulk materials} \\              
        Z_0\frac{|\sigma(\omega)|^2}{\Re \sigma(\omega)}  & \text{2D materials},
    \end{dcases} \label{eq:foms} 
\end{equation}	
where the impedance of free space, $Z_0$, is included in the 2D-material FOM to make it dimensionless. This material metric determines the maximal single-frequency response for many applications, including absorption, scattering, and local density of states~\cite{Miller2016,Miller2014,Miller2017a}, cross density of states~\cite{Shim2019}, near-field radiative heat transfer~\cite{Miller2015}, high-radiative-efficiency plasmonics~\cite{Yang2017}, free-electron radiation~\cite{Yang2018}, Raman scattering~\cite{Michon2019}, and more. \Eqref{foms} is for scalar, nonmagnetic materials; more generally, with a tensor susceptibility $\vect{\chi}$ or conductivity $\vect{\sigma}$, the material metric becomes $\left\| \vect{\chi}^\dagger \left(\Im \vect{\chi}\right)^{-1} \vect{\chi} \right\|_2$ (\citeasnoun{Miller2016}) and $Z_0 \left\| \vect{\sigma}^\dagger \left(\Im \vect{\sigma}\right)^{-1} \vect{\sigma} \right\|_2$ (\citeasnoun{Miller2017}), respectively, where $\left\|\cdot\right\|_2$ denotes the matrix 2-norm~\cite{Trefethen1997}.

\begin{figure*} [tb!]
\centering\includegraphics[width=1\linewidth]{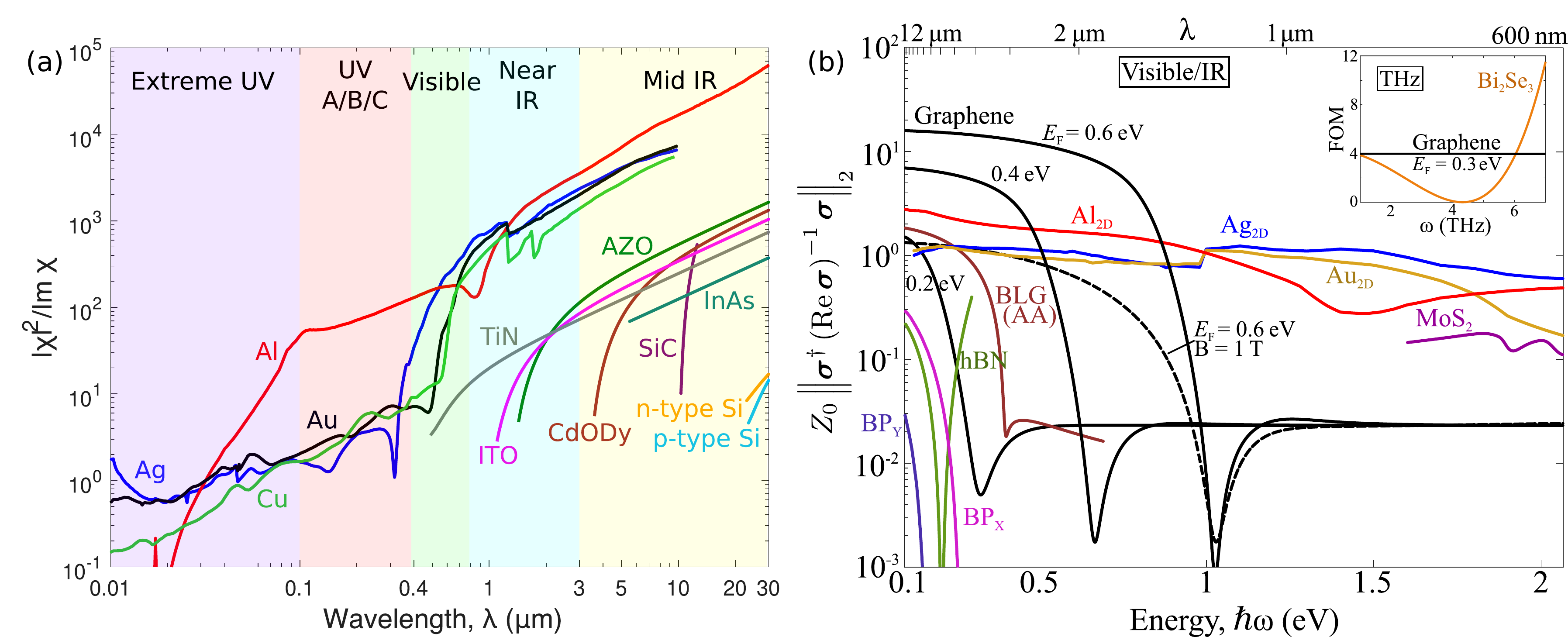} 
\caption{(a) Comparison of representative bulk, polaritonic (and/or lossy) materials via the material figure of merit ${|\chi(\omega)|^2}/ \Im{\chi(\omega)}$. Conventional metals (Ag, Au, Al, Cu)~\cite{Palik1998} outperform alternative plasmonic materials (aluminum-doped ZnO (AZO)~\cite{Naik2013}, Dysprosium-doped CdO (CdO:Dy)~\cite{Sachet2015}, SiC~\cite{Francoeur2010}, TiN~\cite{Hibbins1998}, ITO~\cite{Franzen2008}, doped InAs~\cite{Law2013}, n-type and p-type Si~\cite{Basu2010}) in the visible and infrared (albeit at the expense of large permittivity real parts). Drude material susceptibilities exhibit linear increases in material FOM as a function of wavelength, explaining the linear trend in the figure. (b) Comparison of various 2D materials by material FOM $Z_0 \left\| \vect{\sigma}^\dagger \left(\Im \vect{\sigma}\right)^{-1} \vect{\sigma} \right\|_2$ (or $Z_0{|\sigma(\omega)|^2}/{\Re \sigma(\omega)}$ in the scalar case). Here, we compare: graphene at different Fermi levels~\cite{Jablan2009} (solid black lines) and magnetic-biasing~\cite{Hanson2008} (dashed black line), AA-stacked bilayer graphene~\cite{Wang2016} (dark red), hBN~\cite{Brar2014} (green), MoS$_{2}$~\cite{Liu2014a} (purple), the anisotropic conductivity components of black phosphorus~\cite{Low2017} (BP, pink and dark purple), and three 2D metals~\cite{GarciaDeAbajo2015}, Al (red), Ag (blue), and Au (gold). High-Fermi-level graphene and 2D silver offer the largest possible responses at infrared and visible wavelengths, respectively. The inset compares graphene at THz frequencies~\cite{Rana2008,Ju2011,Low2014} to the topological insulator Bi$_{2}$Se$_{3}$~\cite{DiPietro2013}, which can have a surprisingly large FOM. For (b), reprinted with permission from Ref.~\cite{Miller2017a}, American Chemical Society.}  
    \label{fig:FOM} 
\end{figure*} 
Intuitively, the material metric of \eqref{foms} makes sense: larger absolute susceptibilities imply the ability to sustain large currents, while the imaginary part of the susceptibility must dampen resonant response. \Figref{FOM}(a,b) compares different polaritonic materials against the material FOMs $f(\omega)$ of \eqref{foms} for experimentally characterized bulk and 2D materials. In the figure for bulk materials, across a spectrum ranging from the extreme UV to the mid-infrared, there is a clear trend for increasing $f(\omega)$ with wavelength, which can be attributed to Drude-like response in such materials. For a Drude susceptibility $\chi(\omega) = -\omega_p^2 / (\omega^2 + i\gamma\omega)$, the material FOM is given by $\omega_p^2 / \gamma\omega$, and is therefore exactly proportional to wavelength. Variations from linear dependence thus represent non-Drude features in the material susceptibilities. The increasing material FOM with wavelength may be compensated by frequency-dependent constants in the response function; for example, the far-field scattering bounds~\cite{Miller2016} multiply the material FOM by a factor $\omega/c$ that exactly compensates a linear increase with wavelength. As we discuss further in \secref{RPP}, the large values of $|\chi|^2 / \Im \chi$ for noble metals at infrared frequencies may represent bounds that are not achievable in practice (due to infeasible synthesis requirements), in which case polar-dielectric materials, transparent conducting oxides, and doped semiconductors may all be viable alternatives. In the case of 2D materials, per \figref{FOM}(b), one can see that graphene with a large Fermi level appears ideal at photon energies below \SI{1}{eV}, while 2D Ag, Al, and Au all perform very well at higher photon energies. In that figure we take the 2D-material limit from bulk properties of the metals; intriguingly, ab-initio calculations suggest that actual single-layer sheets of 2D materials may have significantly larger material FOM than their infinitely thin bulk counterparts~\cite{Sundararaman2018}.

The bulk-material figure of merit of \eqref{foms} appears in other contexts as well. In \citeasnoun{Arbabi2014}, it is shown that $|\chi|^2 / \Im \chi$ is the key material metric determining a geometry-independent fundamental limit to propagation length in plasmonic waveguides. The essence of that derivation similarly approaches the problem as one of identifying the maximum possible induced polarization currents. (It has been suggested~\cite{Raether1988,Barnes2003,Dionne2005,Barnes2006,Homola2006} that the propagation length of a surface plasmon on a planar interface follows an expression that ultimately is proportional to $(\Re \varepsilon)^2 / \Im \varepsilon$, which is very similar to the material metric. However, this is derived with a Taylor expansion that is invalid on resonance, as explained in Appendix E of Ref.~\cite{Miller2016}; the correct expression is in fact proportional to the $\sqrt{\Im \chi}$, which \emph{decreases} as loss decreases, because of the concomitant reduction in group velocity. But these arguments only apply to planar surfaces and patterned surfaces may more closely approach the bound of \citeasnoun{Arbabi2014}.) The inverse of the material metric has also been identified as the fundamental loss quantity to be minimized in metamaterial-based models~\cite{Tassin2012}.

Epsilon-near-zero materials~\cite{Liberal2017,Kinsey2019} exhibit intriguing phenomena such as distorted channels~\cite{Silveirinha2006,Edwards2008}, high-directivity emission~\cite{Enoch2002,Pacheco2014}, and arbitrarily large phase velocities~\cite{Moitra2013,Li2015}, and offer the possibility for significant enhancement of nonlinear optical response~\cite{Capretti2015,Alam2016,Caspani2016}. But they do not offer any particular benefit for large linear response. At the frequency where the real part of the permittivity crosses zero, the material FOM simplifies to $\left[1 + (\Im \chi)\right]^2 / \Im \chi$, which will tend to be significantly smaller than many of the values in \figref{FOM}(a), due to the modest magnitude of $\chi$.

An ideal polaritonic material has a purely real, negative permittivity with zero loss, in which case the material FOM diverges. A real, negative permittivity over a nonzero, finite frequency band is compatible with causality requirements (e.g., the Kramers--Kronig relations), and in theory ultra-low loss metals may be achievable in materials with artificially large lattice spacing or in designer organometallic compounds~\cite{Khurgin2010}. The bounds for any physical power-flow quantity cannot themselves diverge, and must be regularized by other effects in the presence of a lossless material. In the case of cross-sections \emph{per volume}, the bounds \emph{can} diverge as nonzero response is possible with arbitrarily small volumes. However, cross-sections themselves cannot diverge as radiative losses must become dominant as absorptive losses go to zero~\cite{gustafsson2019upper,Kuang2020subm,molesky2020}. In the near field, there is not necessarily any radiative coupling; one form of regularization would be the breakdown of a local bulk susceptibility at the large wavevectors that are accessible with very small loss. In either scenario, lower-loss materials towards such regularizations would represent improved response relative to the current state-of-the-art.
From a microscopic perspective, two fundamental sources of loss are intrinsic quantum linewidth broadening~\cite{Kuzyk2006} as well as inhomogeneous broadening, but typical metals exhibit higher losses than required by these sources and thus significant improvements may be possible~\cite{Crowell2020}.

\begin{figure*} [tb!]
\centering\includegraphics[width=1\linewidth]{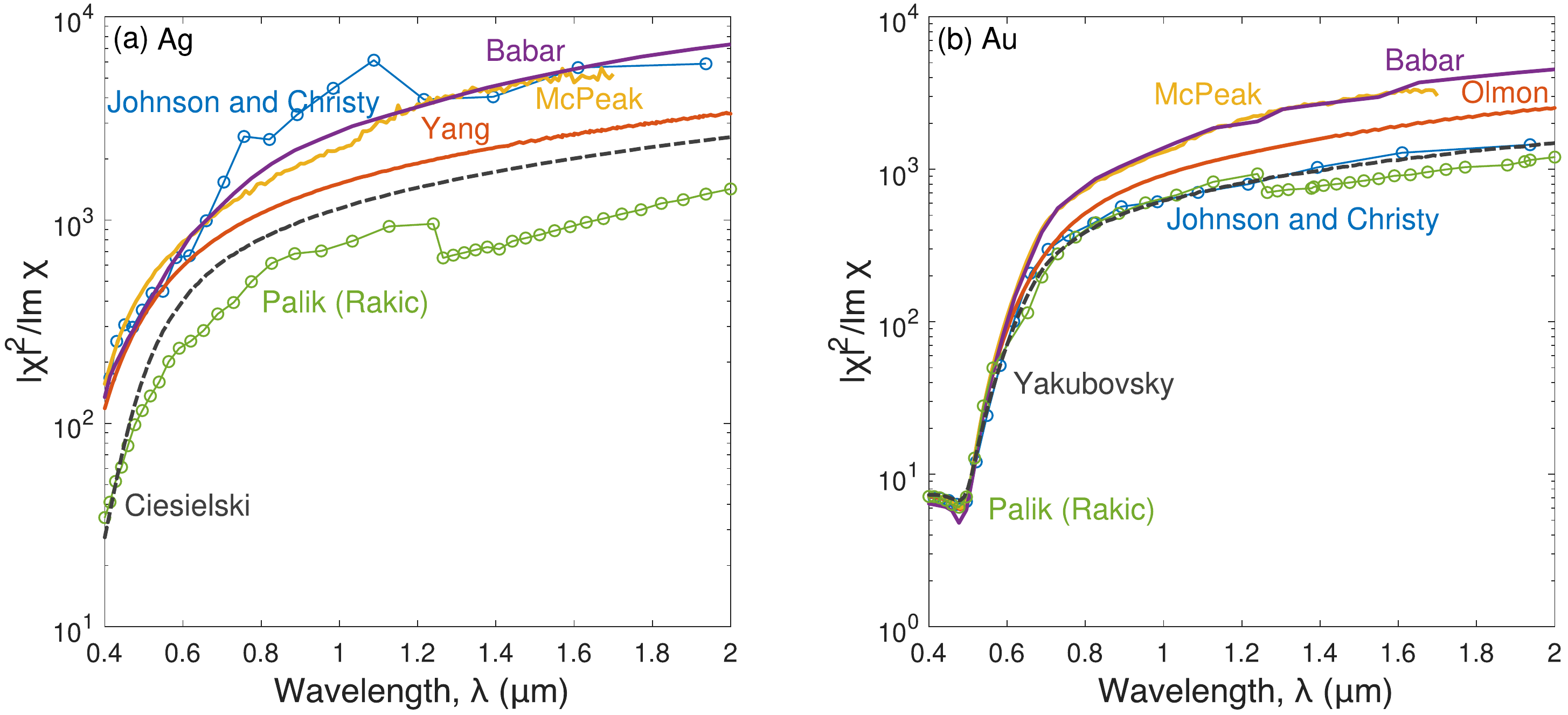} 
    \caption{Figure of merit, $ f(\omega) = {|\chi(\omega)|^2}/{\Im{\chi(\omega)}}$, of different experimental data~\cite{Johnson1972,Palik1998,Olmon2012,Babar2015,McPeak2015,Yang2015a,Yakubovsky2017,Ciesielski2017} for (a) Ag and (b) Au. The metric can vary dramatically even within the same material, depending on the synthesis techniques and experimental conditions. The names represent the authors from whom the data was obtained, and all of the curves are for bulk, thick films except for the dashed black lines, which are for thin films.}  
    \label{fig:agauFOM}
\end{figure*} 

There has been significant effort in mitigating losses in polaritonic materials~\cite{Khurgin2010,West2010,Boltasseva2011,Khurgin2012,Naik2013,Khurgin2015}. The synthesis techniques for a given material can dramatically affect the material FOM, as depicted in \figref{agauFOM}. High-quality films of Ag and Au can be deposited using focused-ion-beam lithography or other techniques~\cite{McPeak2015}. However, it is shown that, up to a threshold, the thickness of the film will adversely increase the optical loss~\cite{Yakubovsky2017,Ciesielski2017}. The dielectric function of bulk silver is sensitive to environmental conditions and influenced by extrinsic effects, such as surface, impurity, and grain boundary scattering~\cite{Yang2015a}. On the contrary, bulk gold is less sensitive to sample morphology and variation between different bulk measurements~\cite{Johnson1972,Palik1998,Babar2015} are mostly likely caused by systematic errors~\cite{Olmon2012}. There are other approaches to mitigating loss as well. One approach is to engineer the free-carrier concentration~\cite{Hoffman2007,Blaber2010,Naik2010,Naik2011,Naik2012,Kim2013,Kim2013a,Kim2014,Agrawal2018}. Another is to use gain media to compensate for loss~\cite{Bergman2003,Gather2010,Stockman2011}. Polar dielectrics supporting surface phonon-polaritons tend to naturally have lower losses and are good polaritonic media at mid-infrared frequencies~\cite{Caldwell2015,Low2017,Caldwell2019}. While all these approaches help reduce loss as measured by the imaginary part of susceptibility, they do not necessarily translate to an enhancement in the material FOM. For example, reducing the free-carrier concentration decreases the magnitude of the real part of susceptibility, in addition to decreasing its imaginary part, thus resulting in smaller material FOM. Thus a useful measure of loss is the inverse of the material FOM, ${\Im{\chi(\omega)}}/{|\chi(\omega)|^2}$. Under this metric, the various approaches mentioned above may not be very effective. Countertuitively, measures that increase the imaginary part of the susceptibility could help reduce absorptive losses.



\begin{figure*} [h!]
\centering\includegraphics[width=1\linewidth]{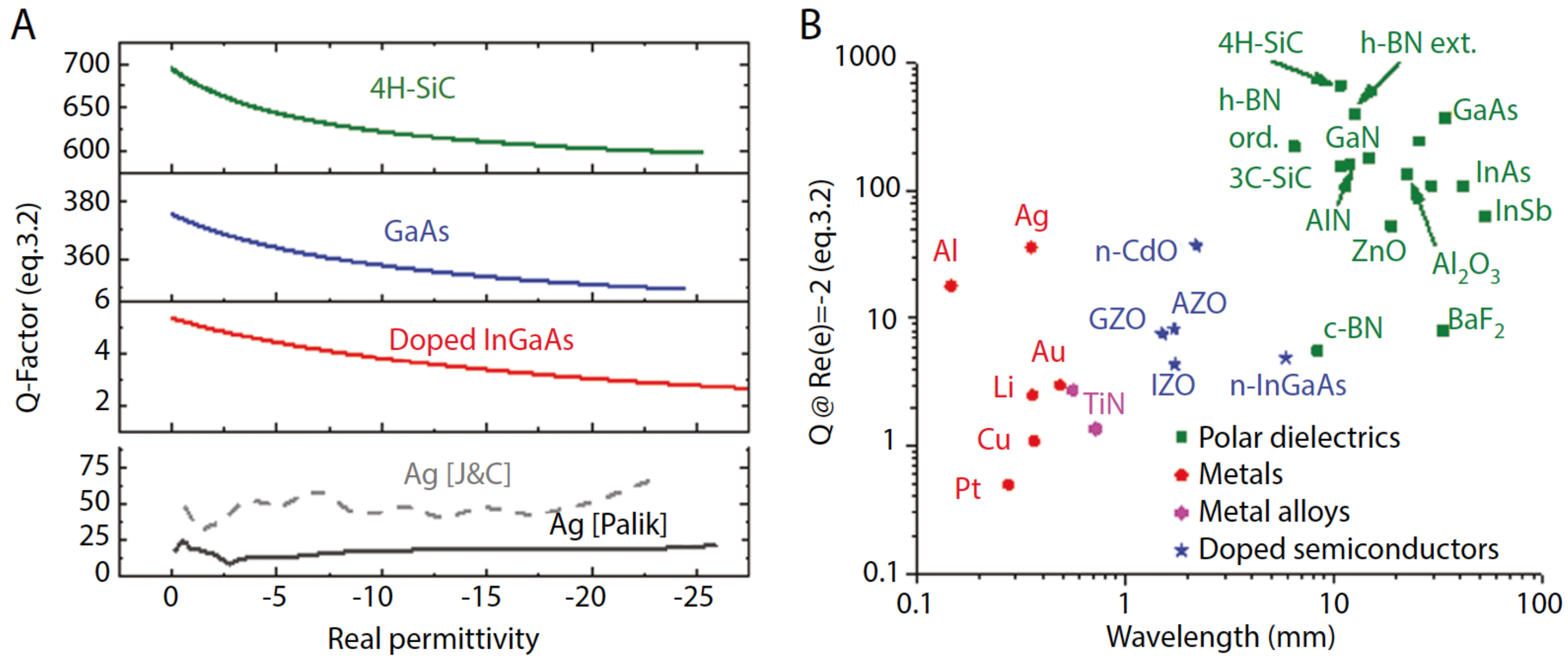} 
    \caption{(a) Quality factor (using \eqref{qfeng},valid under quasistatic approximation) for representative materials as a function of real part of permittivity. For Ag, optical constants were obtained from both Palik~\cite{Palik1998} and Johnson and Christy (J$\&$C)~\cite{Johnson1972}. (b) Comparison of quality factor (\eqref{qfeng}) for different materials, ranging from polaritonic materials---metals, metal alloys, and doped semiconductors---to polar dielectrics. Reprinted with permission from Ref.~\cite{Caldwell2015}, De Gruyter.}  
    \label{fig:qfactor} 
\end{figure*} 
\subsection{Quality factor}
\label{sec:QF}
Quality factor is another measure of loss, indicative of both the relative loss rate per cycle in a resonator as well as the linewidth of the scattering contribution from a single resonance. In general, it is a function of both the resonator geometry and its material properties, but for high-confinement, highly subwavelength polaritonic resonances, the quasistatic nature of the resonances implies that material loss is the dominant source of loss and is geometry independent.

Using simple integral relations for quasistatic fields, \citeasnoun{Wang2006} derived the quality factor of any low-loss polaritonic resonator:
\begin{align}
    Q = \frac{\omega}{2}\frac{\frac{\partial}{\partial{\omega}}{\left[\Re{\chi(\omega)}\right]}}{{\Im{\chi(\omega)}}}. \label{eq:qfeng}
\end{align}
For Drude materials with loss rate $\gamma$, the $Q$-factor expression simplifies to $Q = \omega/\gamma$. On physical grounds, Khurgin~\cite{Khurgin2012} has pointed out that quasistatic resonances have vanishingly small magnetic-field energies relative to their electric-field energies. The magnetic-field energy serves the role of ``kinetic energy'' for the resonator, and is replaced by the kinetic inductance of the free electrons. Energy stored in the free electrons, however, will dissipate at a rate proportional to $\gamma$, independent of the geometrical details of the structure.

Beyond the low-loss regime, \citeasnoun{Raman2013} extends the result of \eqref{qfeng} to any loss regime for Drude-Lorentz-oscillator material models. For Drude-Lorentz oscillators with damping rates $\Gamma_n$, they define a variable $\gamma_{\rm max}(\omega)$ that is a frequency-dependent weighted average of $1/2$ times the $\Gamma_n$ rates, $\gamma_{\rm max}(\omega) = \sum_n \theta_n(\omega) \Gamma_n/2$, where $\theta$ is a spectral weighting factor~\cite{Raman2013}. Then the quality factor is bounded below by a simple ratio of the resonant frequency $\omega$ to $\gamma_{\rm max}$:
\begin{align}
Q \geq \frac{\omega}{2\gamma_{\textrm{max}}(\omega)}, \label{eq:qraman} 
\end{align}
where the extra factor of 2 in the denominator arises from the definition of $\gamma_{\rm max}$ as a weighted average of \emph{half} of the Drude loss rates. Technically, \eqref{qraman} does not require a quasistatic approximation, but it does require material losses to dominate relative to radiative losses.

These derivations can be adapted easily to 2D materials, by making the replacement $\omega\chi(\omega) \rightarrow i\delta_S(\xv) \sigma(\omega)$, where $\delta_S(\xv)$ is a delta function on the surface of the (not necessarily planar~\cite{Asmar2015}) 2D material. The analog of \eqref{qfeng} for a 2D material is:
\begin{align}
    Q = -\frac{\omega^2}{2}\frac{\frac{\partial} {\partial{\omega}}\left[\frac{{\Im{\sigma(\omega)}}}{\omega}\right]}{{\Re{\sigma(\omega)}}}. \label{eq:qfeng2d}
\end{align}
Beyond the quasistatic approximation, it \emph{is} possible to narrow the linewidth beyond the expressions of \eqrefrange{qfeng}{qfeng2d}, through e.g. Fano resonances~\cite{Fano1961,Limonov2017} and near-field coupling~\cite{Stockman1996,Kottmann2001}, but such effects must necessarily occur at larger size scales, without the highly subwavelength confinement available in the quasistatic limit.

\Figref{qfactor} plots the quality factor computed by \eqref{qfeng} for a wide variety materials. To compare loss rates at very different frequencies, in \figref{qfactor}(a) the $Q$ factor of representative materials is shown over a range of moderate values of the real parts of their permittivities. \Figref{qfactor}(b) shows the $Q$ factors of many materials at the frequencies for which $\Re \varepsilon = -2$, where a subwavelength sphere exhibits a surface-plasmon or surface-phonon resonance. The $Q$ factor near these resonances can be expressed as $\omega/\gamma$ where $\omega$ refers to the resonance frequency. The $Q$ factor tends to increase with wavelength because optical phonon lifetimes for polar dielectrics are typically orders of magnitude larger than those of plasmonic metals, which more than compensates for the (roughly an order-of-magnitude) reduction in resonance frequency. Doped semiconductors are intermediate between the two classes, with their loss rates several times smaller, but resonance frequencies slightly smaller, than their metal counterparts.

\subsection{Real part of permittivity}
\label{sec:RPP}
Polaritonics in the mid-IR spectrum is rich with applications in sensing and selective thermal emission~\cite{Greffet2002,Mason2011,Liu2011}, given that a wide variety of molecules exhibit fundamental vibrational and rotational modes in the mid-IR and that blackbody emission peaks in this range for typical temperatures~\cite{Law2012,Caldwell2015,Zhong2015}. As discussed above, metals can have very large material FOM $f(\omega) = |\chi|^2 / \Im \chi$ at such frequencies because the real parts of their permittivities tend to scale as $1/\omega^2$. However, achieving the corresponding scattering bounds may be unrealistic. Achieving polaritonic resonances in materials with large negative real permittivities may require difficult-to-fabricate feature sizes. As an example, for ellipsoids to have a quasistatic /plasmonic resonance at a particular frequency requires their depolarization factors $L$ to equal $\Re(-1/\chi)$ (\citeasnoun{Miller2016}), which requires increasingly large aspect ratios as $L = \Re(-1/\chi) \rightarrow 0$. As an example, for a resonance at $\SI{5}{\mu m}$ wavelength in silver would require an aspect ratio greater than 50, whereas the highest aspect ratios fabricated to date are roughly 30 (\citeasnoun{Nanorods2003}). It is possible to shift resonances to longer wavelengths without high aspect ratios by increasing their size, and correspondingly the radiative damping, but the confinement is reduced and the materials start behaving more like perfect conductors rather than plasmonic materials~\cite{Law2013}. Large, negative real part of permittivities are also undesirable for certain transformation-optics based devices and applications. For instance, Ref.~\cite{Cai2007} has designed a non-magnetic, cylindrical cloak at optical frequencies that require the real part of permittivity of the constituent metal wires to have a similar magnitude to that of the surrounding dielectric.

\begin{figure*} [h!]
\centering\includegraphics[width=0.6\linewidth]{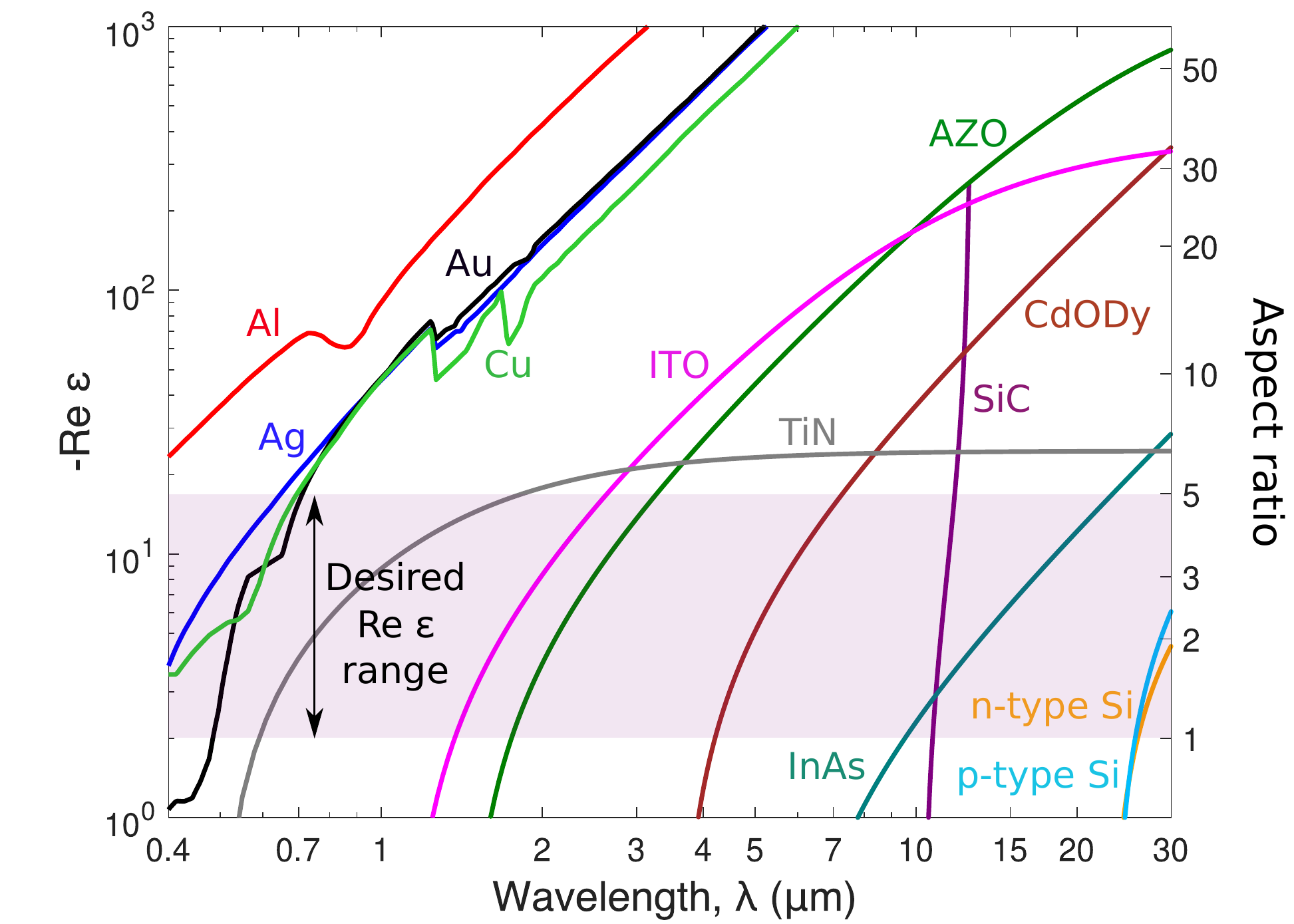} 
    \caption{Comparison of (negative) $\Re \varepsilon$ of various polaritonic materials for a wide range of frequencies, ranging from visible to mid-IR. The pink shaded region shows the allowed range of $\Re \varepsilon$ for realistic aspect ratio of 1 to 5 for a prolate spheroid. Given this range, conventional metals are ideal in the visible, whereas alternative plasmonic materials are suited for infrared spectrum. The extent to which the real part of permittivity of different materials responds to wavelength varies, with SiC demonstrating very small wavelength range over the pink shaded region and doped InAs the largest in the mid-IR.}  
    \label{fig:FOMreal} 
\end{figure*} 

\Figref{FOMreal} compares the magnitude of $\Re \varepsilon$ for various polaritonic materials in the visible and infrared spectrum. The shaded region corresponds to permittivity values whose real parts are between -2 and -16.9, which for nanorod-like ellipsoidal nanoparticles maps to aspect ratios from 1 to 5. This may be considered the region within which true quasistatic resonances can be achieved.

A subtlety that arises for high-loss materials is that the negativity of the real part of the permittivity no longer becomes the true delineation of whether a material supports quasistatic polaritonic resonances. In quasistatic electromagnetism, Maxwell's equations simplify to Poisson's equation for the quasistatic fields. Poisson's equation can be transformed to a surface-integral equation~\cite{Fuchs1975,Ouyang1989,GarciadeAbajo2002, Mayergoyz2005,Kellogg1929,Miller2014} for surface-charge configurations $\sigma$ at all material interfaces, a surface-integral equation that can be written in the form $\hat{K} \sigma - \Lambda \sigma = -s$, where $\hat{K}$ is the Neumann-Poincare operator~\cite{Mayergoyz2005,Kellogg1929,Sandu2013} that is a Green's-function convolution operator, $\Lambda$ depends on the material susceptibility $\chi$ (assuming vacuum exterior, easily generalizable) via $\Lambda = 1/2 + 1/\chi$, and $s$ is a source term proportional to the incident field. The key aspect relevant to this discussion is that in an eigendecomposition of $\hat{K}$, the response will be maximal if the real part of $-1/\chi$ is in the range $[0,1]$. This is the true condition for plasmonic-like response for a lossy material: $\Re(-1/\chi) \in [0,1]$, generalizing the simple negative-permittivity condition for materials with nontrivial loss rates.

\subsection{High-radiative-efficiency plasmonics}
\label{sec:radeff}
In the preceding sections, it was emphasized that one wants quasistatic resonances for maximum confinement in polaritonic response, and that coupling to radiative channels reduces such response. Yet for applications where high radiative efficiency is important, in areas such as far-field imaging~\cite{tam2007plasmonic,Anker2009b}, photovoltaics~\cite{atwater2010plasmonics,brongersma2014light}, and quantum nanophotonics~\cite{de2012quantum}, it is important to identify bounds that incorporate radiative-efficiency constraints. This problem is considered in \citeasnoun{Yang2017}, where it is shown that for both far-field scattering and near-field LDOS quantities, one can derive bounds with the additional constraint of high radiative efficiencies. In particular, the results of \citeasnoun{Yang2017} show that imposing a minimum radiative efficiency $\eta_{\rm min}$ (for hard-to-achieve radiative efficiencies above 50\%) effectively reduces the maximum possible scattering response by a factor $\eta_{\rm min}\left(1 - \eta_{\rm min}\right)/4$, in which case one can define the material FOM by
\begin{align}
    f(\omega) = \frac{\eta_{\rm min}\left(1 - \eta_{\rm min}\right)}{4} \frac{|\chi(\omega)|^2}{\Im \chi(\omega)}.
    \label{eq:radeff}
\end{align}
The new material FOM, given by \eqref{radeff}, explicitly identifies the tradeoff in response that must be sacrificed to achieve high radiative efficiency. 

\begin{figure*} [tb!]
    \centering\includegraphics[width=1\linewidth]{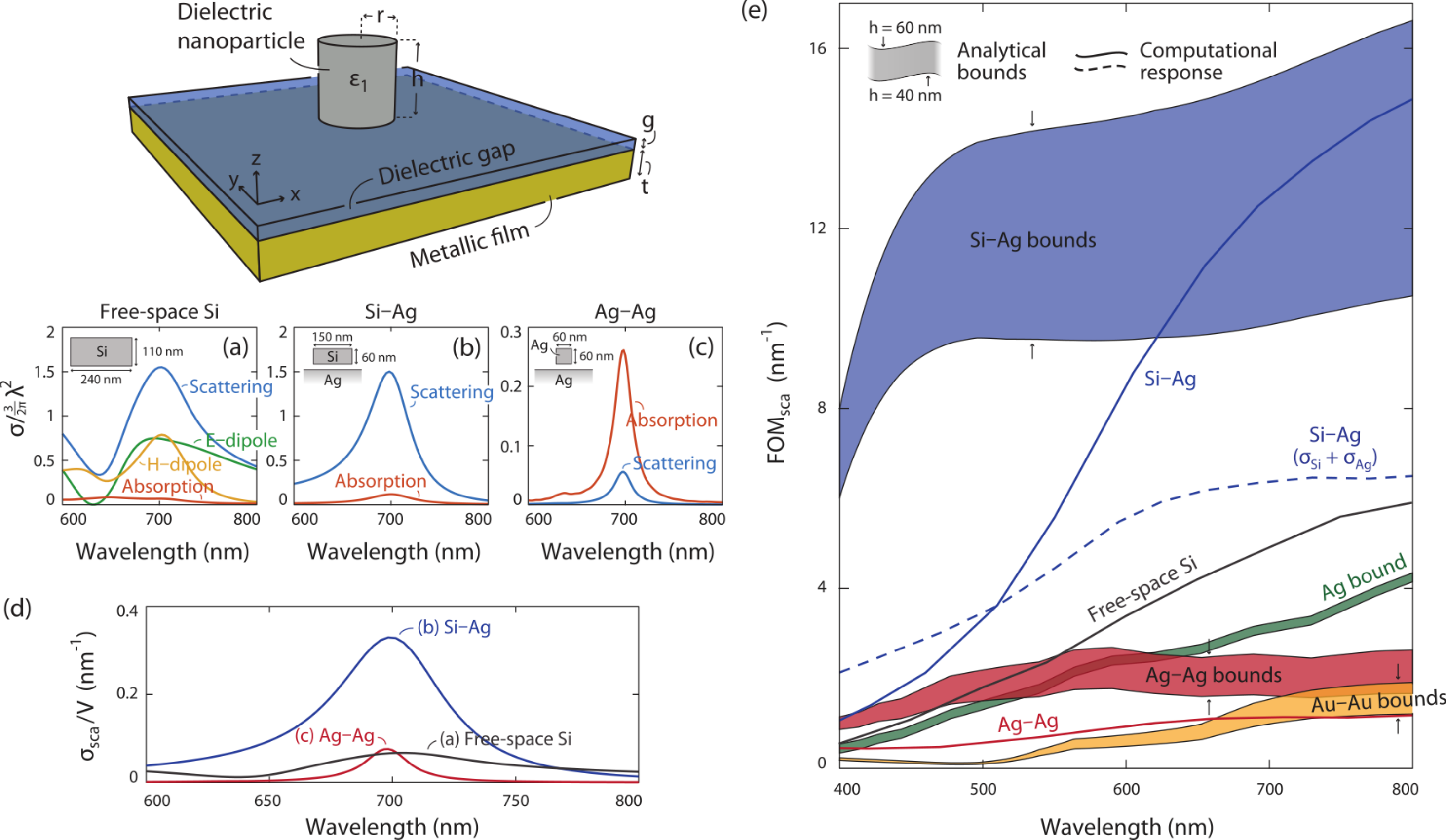}
    \caption{Schematic of hybrid dielectric--metal resonator, as shown in the top left. Scattering and absorption cross sections of (a) Si cylinder in free-space and (b,c) Si and Ag cylinders, respectively, above a semi-infinite Ag substrate with gap thickness $g =$ 2 nm. Geometrical parameters (insets) are chosen to align their resonant wavelengths at 700 nm. The three structures are all illuminated by normally incident plane waves. In (b,c), the absorption includes the dissipation in both the particle and the substrate. (d) The dielectric–metal structure shows the highest per-volume scattering cross-section, because it simultaneously achieves large scattering cross-section $\sigma_{\textrm{sca}}$, high radiative efficiency $\eta$, and small particle volume $V$. (e) In the visible regime, the scattering capabilities of metal–metal geometries (Ag--Ag and Au--Au bounds), free-space metallic (Ag bound), and free-space dielectric (Si free-space) scatterers all fall short when compared with the dielectric--metal (Si--Ag) scatterer, which also approaches its own upper bound. For the Si--Ag and Ag--Ag structures, the gap size is fixed at 5 nm; the cylinder (both Si and Ag) height $h$ ranges from 40 to 60 nm in order to tune the resonant wavelength. Reprinted with permission from Ref.~\cite{Yang2017}, American Chemical Society.}  
    \label{fig:hre} 
\end{figure*} 

The radiative-efficiency-constrained bounds enable comparison~\cite{Yang2017} of all-metal~\cite{esteban2010optical,moreau2012controlled,belacel2013controlling,rose2014control,faggiani2015quenching} and hybrid metal--dielectric~\cite{Yang2017,devilez2010compact,rusak2014hybrid} approaches to high-radiative-efficiency plasmonics. Shown in \figref{hre} is one set of findings from \citeasnoun{Yang2017}: a hybrid silicon-on-silver resonator could have superior plasmonic properties to an all-silver resonator when radiative-efficiency constraints are included. Shown in \figref{hre}(a-d) are the absorption and scattering cross-sections of silicon-only, silicon--silver, and silver-only optimized resonator designs, with the hybrid silicon-on-silver resonator showing the largest scattering cross-sections per volume. Moreover, as shown in \figref{hre}(e), these resonators approach their scattering-efficiency bounds, including the metric of \eqref{radeff}. Also shown in \figref{hre}(e) are the bounds for silver-only and gold-only nanoresonators, which lie below the actual scattering performance of the designed silicon-on-silver structure (solid blue lines). This implies that there is no silver-only or gold-only approach that can ever do better than the designed silicon-on-silver structure, no matter how optimized the patterning is. This shows the power of such bounds: they convey the ability to survey a research field and rank-order certain approaches relative to each other. As shown in \citeasnoun{Yang2017}, the hybrid structures are also superior to metal-only structures for near-field spontaneous-emission enhancements at high radiative efficiency.

\subsection{Nonlocal effects}
\label{sec:NL}
Another important consideration is the effect of \emph{nonlocality} in material susceptibilities for media synthesized at single-nanometer length scales, whereby the polarization currents induced at a point $\vect{x}$ are related to the electromagnetic fields at another point $\vect{x}'$. Such effects both shift resonant frequencies~\cite{DeAbajo2008,Ciraci2012,Mortensen2014,Raza2015} as well as dampen the maximal possible response~\cite{Yan2012,Correas-Serrano2015,Miller2017}. Reference~\cite{Miller2017} considers the maximum response when the material susceptibility is described in a hydrodynamic framework, where the currents behave like fluids with a diffusion constant $D$ and convection constant $\beta$ (both real-valued), in which case the current is the solution of a convection--diffusion equation driven by the electric fields. In a quasistatic framework that is relevant at the length scales where such effects are important, one can show~\cite{Miller2017} that the cross-section bounds depend on a competition of two ``rates:'' the material FOM $f(\omega)$, and a second term relating the size of the scatterer to the ``diffusion'' length in the material. If one defines a radius $r$ of the smallest bounding sphere containing the scatterer, and a plasmonic diffusion length $\ell_D = \sqrt{cD/\omega_p^2}$, for plasma frequency $\omega_p$, the maximum extinction cross-section per area of a 2D-material scatterer is given by~\cite{Miller2017}
\begin{align}
    \frac{\sigma_{\rm ext}}{A} \leq \left[ \left(Z_0 \frac{\left|\sigma_{\rm loc}\right|^2}{\Re \sigma_{\rm loc}}\right)^{-1} + \left( \frac{r^2}{\ell_D^2}\right)^{-1} \right]^{-1},
    \label{eq:nonlocal2D}
\end{align}
where $\sigma_{\rm loc}$ is the local contribution to the conductivity. \Eqref{nonlocal2D} shows a dramatic reduction in response at size scales well below $\ell_D$, serving both to illuminate the effects of nonlocality as well as potentially serving as a means to extract the value of $\ell_D$ itself for any 2D material from experimental measurements. The analog of \eqref{nonlocal2D} for bulk materials is straightforward, replacing the first term with the material FOM $f(\omega)$ from \eqref{foms} and the second term from a volume integral instead of a surface integral. An intriguing next step would be to consider the bounds that are possible in ab-initio material models.

\section{High-index dielectric materials}
\label{sec:dielectric}
Dielectric materials, with $\Re \varepsilon > 0$, can have very small material loss rates, and their dominant loss channel is typically radiation. The positive real part of the permittivity prevents such materials from supporting quasistatic, highly subwavelength resonances (though very small mode volumes are possible~\cite{Robinson2005,Liang2013,Hu2016,Choi2017}), and it elevates the importance of geometrical patterning in their response. Independent of the patterning, however, there is an important geometry-independent sum rule governing their response and identifying high refractive index as a key metric for dielectric materials.

In a homogeneous medium with refractive index $n$, the density of plane-wave electromagnetic states is proportional to $n^3$ and $\omega^2$ at frequency $\omega$ (\citeasnoun{Yariv1989}). If we consider an electric dipole radiating at a any point in the medium, it will efficiently couple to half of these states (effectively the half with the same polarization). The power that it radiates will be directly proportional to the density of states, and is called the (electric) \emph{local} density of states, LDOS~\cite{Novotny2012}. The LDOS is a measure of the relative energy density of modes at a given point in space relative to the total density of modes~\cite{OskooiJo13-sources}. Denoting the free-space electric LDOS by $\rho_0$, it is given by the expression~\cite{Joulain2003}
\begin{align}
    \rho_0(\omega) = \frac{n^3\omega^2}{2\pi^2c^3}.
    \label{eq:rho0}
\end{align}
Now consider a dipolar source in a \emph{structured} medium of refractive index $n$. For a point source at some position $\vect{x}$, the LDOS $\rho(\vect{x},\omega)$ is highly dependent on the position $\vect{x}$, the frequency $\omega$, and the material and structuring. But, if one considers a certain integral over all frequencies (and discard a near-field coupling term that typically, though not always, corresponds to emission into material-absorption pathways), Barnett and Loudon showed~\cite{Barnett1996} that there is a simplifying sum rule from causality arguments. They considered a source point in vacuum, with a structured medium surrounding it, and showed that the integral over all frequencies of the relative difference between the structured-medium LDOS and the vacuum LDOS, i.e. $\left(\rho - \rho_{\rm vac}\right)/\rho_{\rm vac}$, is precisely 0. We can generalize their argument slightly, pointing out that if the source is in a medium with constant refractive index $n$ (over all frequencies), the same sum rule should apply to the structured-medium LDOS relative to the \emph{background} LDOS $\rho_0$ for a homogeneous medium of refractive index $n$, as given in \eqref{rho0}. Then the generalization of the sum rule from \citeasnoun{Barnett1996} is
\begin{align}
    \int_0^\infty \frac{\rho(\vect{x},\omega) - \rho_0(\omega)}{\rho_0(\omega)} \,{\rm d}\omega = 0.
    \label{eq:sumrule}
\end{align}
The sum rule of \eqref{sumrule} states that for any patterning, the total LDOS of a system cannot be modified, on average, over all frequencies. The LDOS can be increased in one frequency range, but thereby must be reduced in another. The only way to increase LDOS over all frequencies is by increasing the refractive index $n$ of the medium itself, which increases $\rho_0(\omega)$ by $n^3$ per \eqref{rho0}. This highlights the key role that refractive index can play in maximizing light--matter interactions.

The role of refractive index to increase the available states has been recognized in a wide variety of systems. Large refractive index enables smaller mode volumes~\cite{Robinson2005,Liang2013,Hu2016,Choi2017} and correspondingly large spontaneous-emission enhancements~\cite{Purcell1995,Agio2013,Sauvan2013}. In a photonic-crystal cavity, a typical defect mode has minimum mode volume $\sim \left(\lambda/2n\right)^3$, in which case larger refractive index widens the bandgap and increases the spatial confinement of the mode~\cite{Yablonovitch1993,Foresi1997,Coccioli1998}. In the quest for miniaturization of nanophotonic ``building blocks,'' plasmonic split-ring resonators tend to exhibit large absorption and can be difficult to fabricate and miniaturize at optical frequencies~\cite{Enkrich2005,Soukoulis2007}. High-index dielectric nanoresonators exhibit strong electric and magnetic resonances~\cite{Zhao2009} and can help overcome such problems~\cite{Kuznetsov2012, Evlyukhin2012}. Large refractive index contrast between the core and air helps to confine the light within the high index medium. As a result, the reduced radiation loss improves the overall $Q$ factors of the resonator~\cite{Kuznetsov2016}. High refractive index is also necessary for the miniaturization of such nanoresonators~\cite{Baranov2017}, since the magnetic resonance of a sphere occurs at a wavelength $\lambda\approx nd$, where $n$ is the refractive index and $d$ is the diameter of a sphere~\cite{Kuznetsov2012, Evlyukhin2012, Staude2013}.  With nanoresonators as building blocks, a high-index material with positive permittivity and unit permeability can create metamaterials with effective permittivities and permeabilities across the four possible quadrants~\cite{Wu2014,Yang2014,Jahani2016}. Nanoresonators with electric- and magnetic-dipole resonances can be used to tailor scattering profiles~\cite{Schuller2007,Ginn2012,Geffrin2012,Fu2013,Person2013,Bakker2015,Vaskin2018,Komar2018}; as one example, effective negative index can be used for Huygens' metasurfaces with no reflection loss and tailored forward scattering~\cite{Decker2015,Liu2018}. Alternatively, with only a single dipolar resonance, one can create perfect electric reflectors with near-unity reflection due to the negligible loss of the dielectric material~\cite{Moitra2014,Moitra2015}. Finally, magnetic reflectors feature maximum electric field at the interface, dramatically enhancing light--matter interaction on that plane~\cite{Liu2014,Esfandyarpour2014}. 

A classical application of the density-of-states enhancement associated with high index is for absorption enhancement in solar cells. Random surface texturing on thick films with plane-wave-like states enables full occupation of all of the internal states, an $n^3$ enhancement relative to external illumination from all angles. The intensity is enhanced in proportion to the product of the density of states with the wave speed, $c/n$, ultimately yielding the $4n^2$ Yablonovitch limit to absorption enhancement~\cite{Yablonovitch1982,Goetzberger1981}. Wavelength-scale pattering has enabled much thinner structures to approach, though generally not surpass, the $4n^2$ limit~\cite{Garnett2010,Yu2010,Han2010,Wang2012,Callahan2012,Martins2013,Ingenito2014,Ganapati2014,Brongersma2014}.

\begin{figure*} [t!]
\centering\includegraphics[width=1\linewidth]{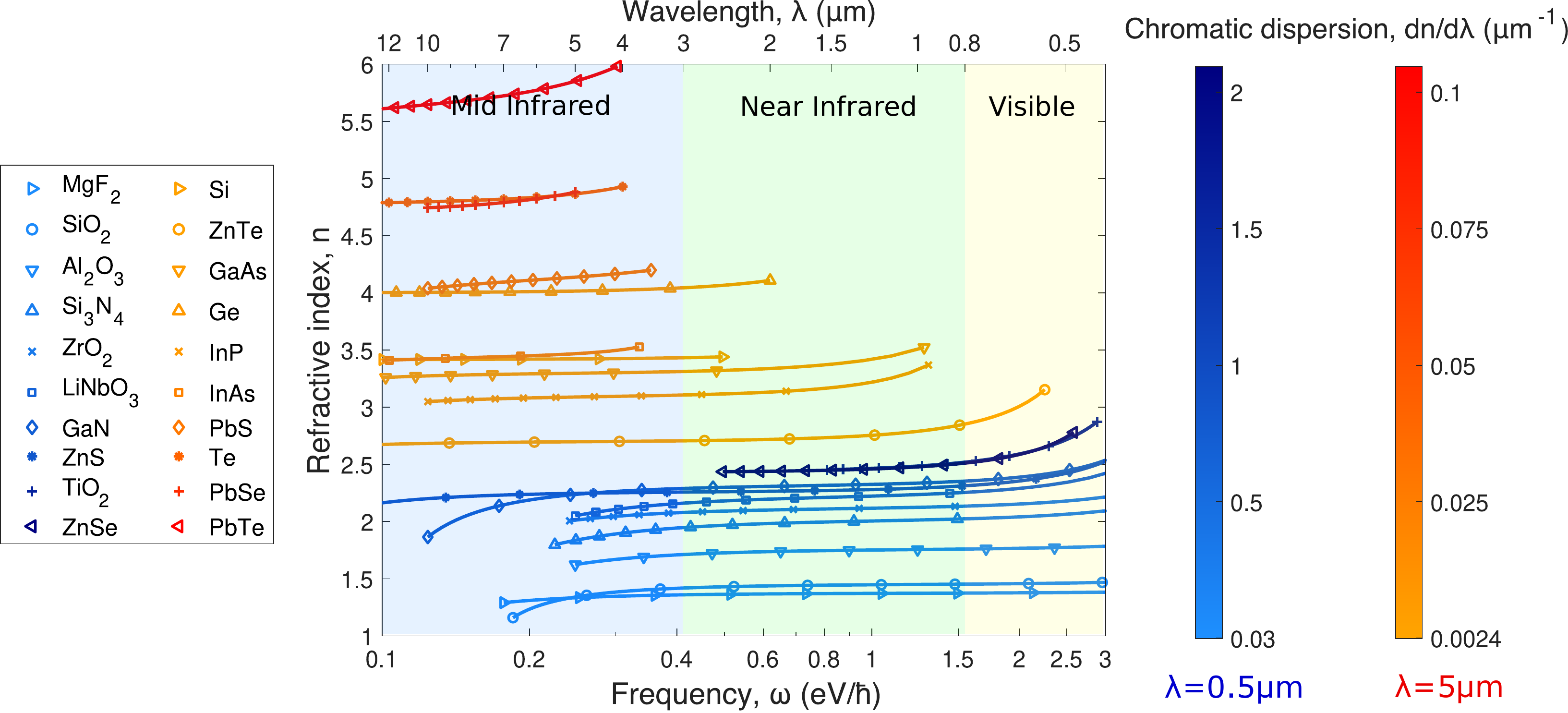} 
    \caption{Comparison of common high-index materials over transparent frequency ranges in the visible and IR spectrum. Curves with blue colors represent materials transparent over the visible and near-IR spectrum (MgF\ts{2}~\cite{Dodge1984}, SiO\ts{2}~\cite{Malitson1965,Tan1998}, Al\ts{2}O\ts{3}~\cite{Dodge1986}, Si\ts{3}N\ts{4}~\cite{Luke2015}, ZrO\ts{2}~\cite{Wood1982}, LiNbO\ts{3}~\cite{Zelmon1997}, GaN~\cite{Barker1973}, ZnS~\cite{Debenham1984,Klein1986}, TiO\ts{2}~\cite{Devore1951}, ZnSe~\cite{Marple1964}), and those with red colors over the mid-IR (Si~\cite{Chandler2005}, ZnTe~\cite{Li1984}, GaAs~\cite{Skauli2003}, Ge~\cite{Burnett2016}, InP~\cite{Pettit1965,Pikhtin1978,Bass2010}, InAs~\cite{Lorimor1965}, PbS~\cite{Zemel1965,Bass2010}, Te~\cite{Bhar1976}, PbSe~\cite{Zemel1965}, PbTe~\cite{Weiting1990}). The chromatic dispersion for the two classes of materials are measured at wavelengths of 500 nm and 5 $\mu$m respectively, with darker shades corresponding to higher values of chromatic dispersion.}  
    \label{fig:highindex} 
\end{figure*} 

\begin{figure*} [t!]
\centering\includegraphics[width=0.5\linewidth]{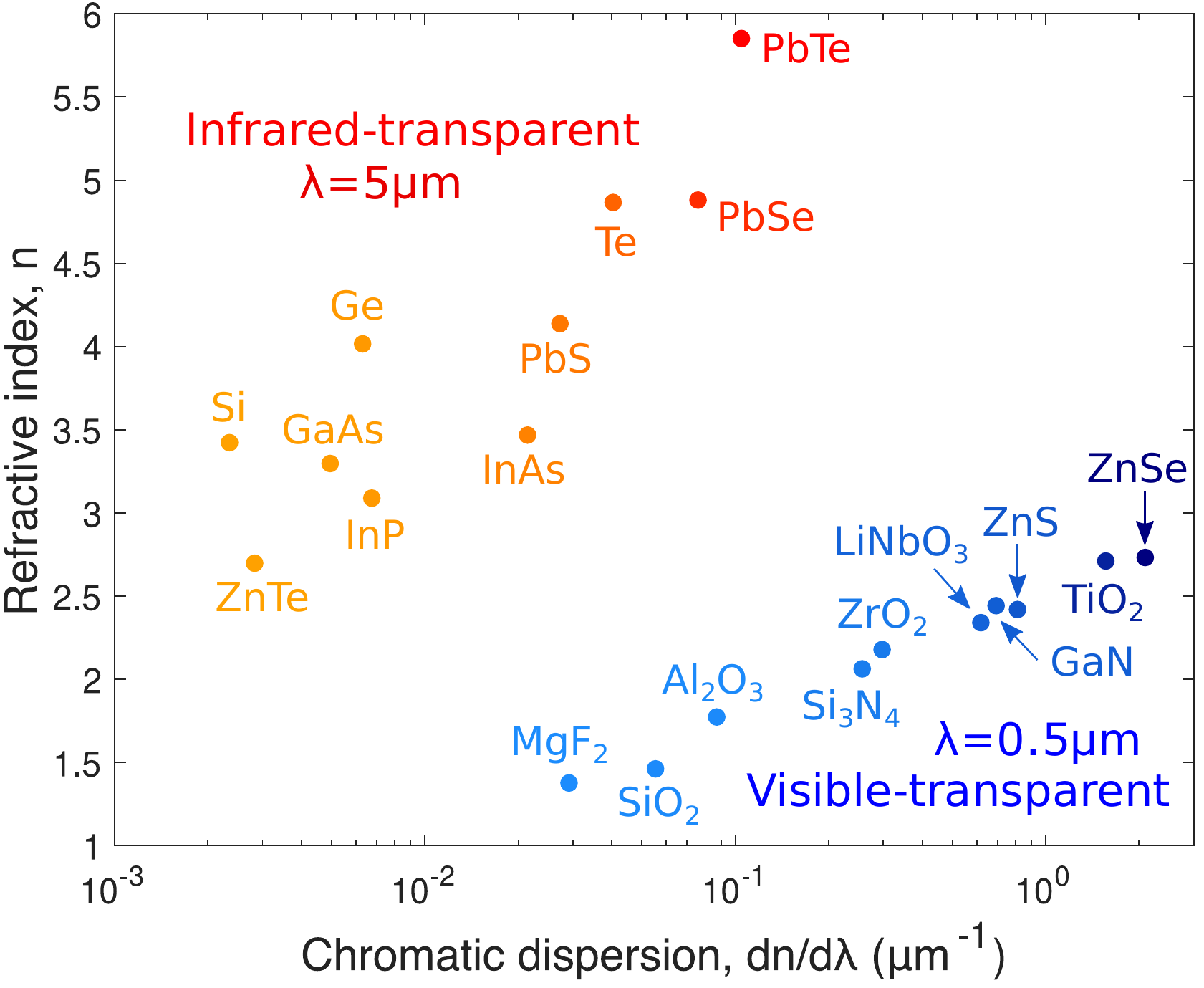} 
    \caption{Dispersion characteristics of high-index materials transparent over the visible and infrared spectrum, as measured by the chromatic dispersion ${\rm d}n / {\rm d} \lambda$ over representative visible and infrared wavelength of 500 nm and 5 $\mu$m respectively. Materials with higher refractive indices tend to be more dispersive.}  
    \label{fig:dispersion} 
\end{figure*} 
Figures \ref{fig:highindex} and \ref{fig:dispersion} show the refractive indices of many common high-index materials over transparency windows at visible and infrared frequencies. Many of the materials that are transparent in the visible and near-IR are polar dielectrics that support phonon polaritons, and thus are not transparent in the mid-infrared. Conversely, many (though not all) of the materials transparent in the mid-IR are not transparent in the visible/near-IR. One can see in both figures that the mid-IR-transparent materials tend to exhibit larger values of $n$ relative to visible/near-IR-transparent materials, consistent with various models showing that the refractive index for semiconductors tends to increase with smaller energy gaps~\cite{Moss1985,Herve1994,Ravindra2007,Tripathy2015,Baranov2017}. One tradeoff that tends to come with higher refractive index is increased chromatic dispersion, as measured by the derivative of the refractive index with respect to wavelength, i.e., $dn/d\lambda$. Figure~\ref{fig:dispersion} clearly shows a nearly linear relationship between refractive index and chromatic dispersion. Thus an important materials-synthesis question is whether higher-index materials with small chromatic dispersion can be synthesized. Such materials would immediately improve the performance of a wide variety of dielectric metasurfaces~\cite{yu2014flat,aieta2015multiwavelength,khorasaninejad2015achromatic,khorasaninejad2016polarization,arbabi2016multiwavelength,kamali2018review,phan2019high,lin2019topology,Shim2019a,Chung2019a,Chung2020}.
            
\section{Comparing Polaritonic and Dielectric Materials}
\label{sec:comp}
The metrics reviewed in Secs.~(\ref{sec:polaritonic},\ref{sec:dielectric}) are not compatible with each other, preventing straightforward comparisons between polaritonic and dielectric media. The bounds in \secref{polaritonic} are inversely proportional to the loss rate of the material, which is practically 0 for many dielectric media. Conversely, the refractive-index sum rule of \secref{dielectric} does not account for the surface waves that are so important in polaritonic media. In this section, we highlight two measures that enable direct comparison of the two systems: total electron number for all-frequency response (\secref{electrons}), and recently developed power--bandwidth metrics for \emph{any} bandwidth of interest (\secref{pb}).

\subsection{Electron number}
\label{sec:electrons}


The total number of electrons $N_e$ in a system is a key property to describe the maximum response of any material. Causality requires any linear electromagnetic response function to be analytic in the upper half of the complex-frequency plane~\cite{Landau1960}, which enables contour-integral-based ``sum rules'' to connect response averaged over all frequencies to certain constants of the scatterer. In quantum systems, this leads to the well-known Thomas--Reiche--Kuhn sum rule (or ``f-sum'' rule)~\cite{Kuhn1925,Thomas1925,Levinger1957,Baxter1992,Wang1999}, which relates the sum of oscillator strengths for energy level transitions to the electron number. One can apply this technique to the extinction cross-section of any optical scatterer, $\sigma_{\rm ext}(\omega)$, yielding the sum rule~\cite{Gordon1963,Yang2015}:
\begin{align}
    \int_0^\infty \sigma_{\rm ext}(\omega) \,{\rm d}\omega = \frac{\pi\omega_p^2}{2c}V = \frac{\pi e^2}{2 \epso m_e c}N_e \approx 1.67 \times 10^{-5} N_e \, [\textrm{m}^2 \textrm{s}^{-1}], \label{eq:fsum} 
\end{align}
where $\omega_p$ is the effective plasma frequency of the material and $V$ is the volume it occupies. \Eqref{fsum} says that the extinction of light summed over all frequencies is determined by the number of electrons in the scatterer, independent of its geometry and the incident-field polarization. In Ref.~\cite{Yang2015}, extinction cross-sections are computed for different canonical geometries and materials including aluminum and silicon, verifying that the sum rule is indeed independent of the nanostructure and material platform. To maximize frequency-integrated extinction, it is advantageous to use materials with large electron number. However, \eqref{fsum} does not provide any guarantees about which frequency ranges will contain the resonances. The sum rule requires susceptibilities that satisfy Kramers--Kronig relations, diminishing to zero at high frequencies. The decay-to-zero requirement, though physically reasonable, means that even ``dielectric'' media (semiconductors, insulators, etc.) have a plasma-like response at large enough frequencies. Such response contributes to integrated extinction, often in a large way due to the negative susceptibility. This obscures the behavior of, for example, a transparent dielectric at optical frequencies, by accounting for transitions that occur at UV and X-ray frequencies and which can be the dominant contributor to the response of \eqref{fsum}.
        
\subsection{Maximal response over nonzero, finite bandwidth}
\label{sec:pb}

\begin{figure*} [t!]
\centering\includegraphics[width=1\linewidth]{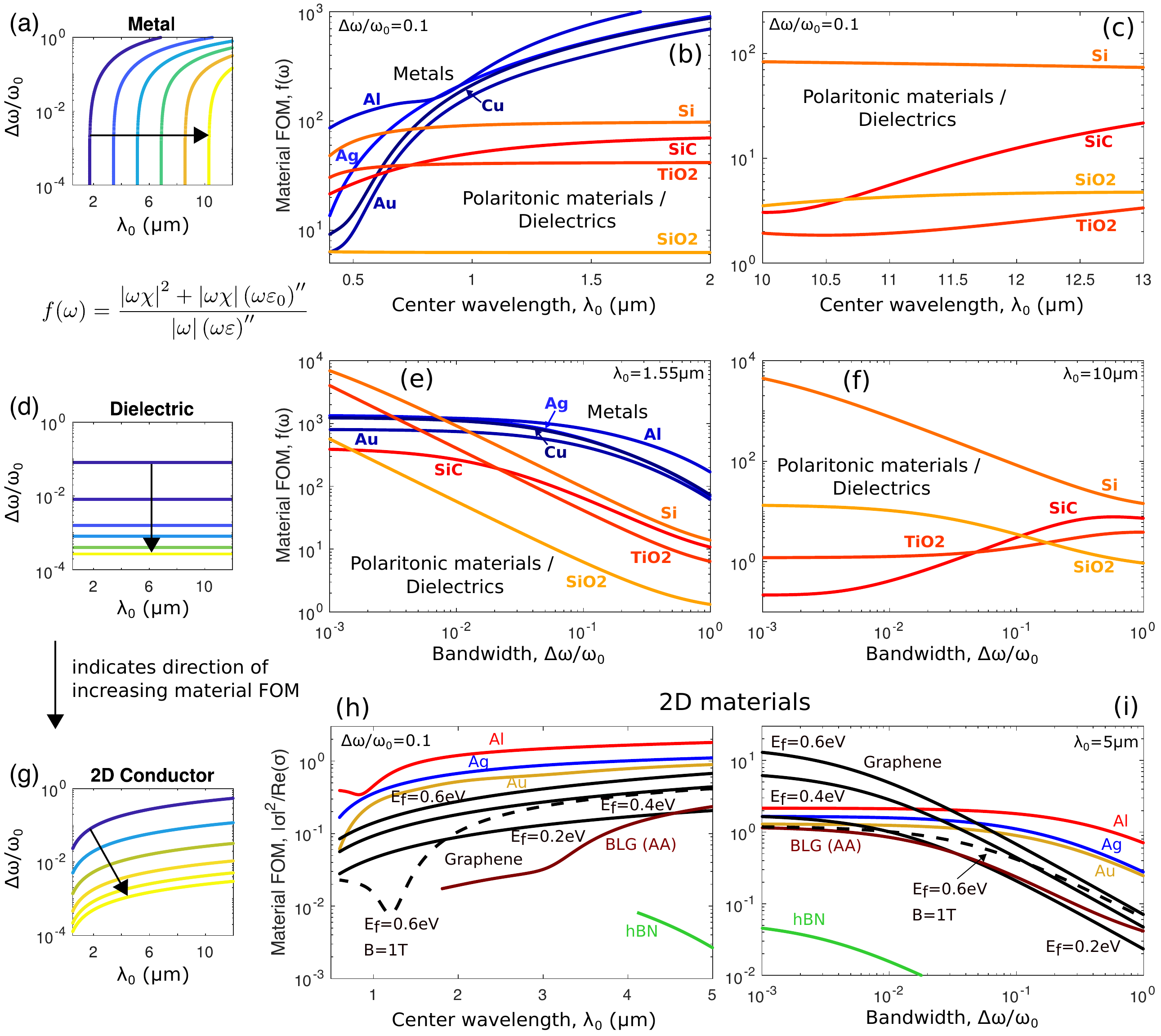} 
    \caption{(a,d,g) Isocurves of material FOM for a Drude metal (with material loss rate $\gamma=0.1\omega_p$), a lossless dielectric (of susceptibility $\chi=9$), and a Drude 2D material (with $\gamma=0.01\omega_p$). The arrows indicate increasing material FOM in each case. (b) Comparison of material FOM for various bulk metals and polaritonic materials (those supporting surface-phonon polaritons) / dielectrics, keeping the bandwidth-to-center-frequency ratio $\Delta \omega / \omega_0$ fixed to 0.1. Part (c) compares surface-phonon-polariton-supporting materials at mid-IR wavelengths. (e),(f) Comparison of material FOM for varying bandwidths relative to the center wavelengths of 1.55 and 10 $\mu$m. At very narrow bandwidths, dielectrics offer greater possible response than metals. (h,i) Comparison of material FOM for 2D materials for different choices of center wavelength and $\Delta \omega / \omega_0$. (2D Al, Ag, and Au properties derived from their bulk counterparts.) Reprinted with permission from Ref.~\cite{Shim2019}, APS.}  
    \label{fig:bandwidthFOM} 
\end{figure*} 

While the Thomas--Reiche--Kuhn sum rule provides information on integrated optical cross section, it does not provide spectral information over any finite bandwidth and often overestimates dielectric-material interactions as described above---the sum rule weighs the entire spectrum equally instead of isolating particular frequencies of interest. To remedy these shortcomings, Ref.~\cite{Shim2019} establishes power--bandwidth limits by combining causality principles, which underlie the sum rules discussed in \secref{dielectric} and \secref{electrons}, with the energy-conservation principles underlying the single-frequency bounds of \secref{polaritonic}. The resulting bounds enable comparisons between polaritonic and dielectric media, and yield the single-frequency and all-frequency bounds in their respective asymptotic limits. The key is to connect the frequency-averaged response over a bandwidth $\Delta\omega$ around a center frequency $\omega_0$ to a single \emph{complex-valued} frequency $\omega = \omega_0 + i\Delta \omega$ (\citeasnoun{Shim2019,Hashemi2012,Liang2013}). (This simple form emerges for a Lorentzian average; other averaging ``windows'' can be used~\cite{Shim2019,Liang2013}.) Then one can derive bounds that depend on material figures of merit for which the material parameters $\chi$ or $\sigma$ are evaluated at this complex frequency, i.e. $\chi(\omega)$ or $\sigma(\omega)$. Then, for nonmagnetic materials with bulk susceptibilities $\chi(\omega)$ or 2D conductivities $\sigma(\omega)$, the material FOM is given by:
\begin{equation}
		f(\omega=\omega_0 + i\Delta\omega) =
    \begin{dcases}
        \frac{\left|\omega\chi\right|^2 + \left|\omega\chi\right| \Delta \omega}{\left|\omega\right| \Im\left(\omega\varepsilon\right)} & \text{3D / bulk materials} \\              
        \frac{|\sigma(\omega)|^2}{\Re \sigma(\omega)}  & \text{2D materials}.
    \end{dcases} \label{eq:f(w)} 
\end{equation}				
(For the general case of anisotropic, magnetic, and even spatially inhomogeneous media, see the Supplemental Material of Ref.~\cite{Shim2019}.) It is evident from \eqref{f(w)} that the material FOM over any nonzero bandwidth yields a finite value even for lossless materials (due to the imaginary part of $\omega$, given by $\Im \omega = \Delta\omega$), and thus enables comparison among all possible optical materials. The FOM for 2D materials has an identical functional form (albeit with a complex frequency) as the single-frequency 2D-material FOM, \eqref{foms}, whereas the functional form of the bulk-material FOM is now slightly more complex in \eqref{f(w)} than its single-frequency counterpart in \eqref{foms}. As in the single-frequency case, the material FOM in \eqref{f(w)} favors large $|\chi(\omega)|$ and small $\Im{\chi(\omega)}$ (and similarly for 2D materials), now evaluated at the complex frequency $\omega=\omega_0 + i\Delta\omega$. One can identify intuitive forms of the material FOMs for small bandwidth ($\Delta\omega \ll \omega_0$), for which the material FOM simplifies to:
\begin{equation}
    f(\omega) \approx
    \begin{dcases}
        \frac{|\chi(\omega)|^2}{\Im{\chi(\omega)}} & \text{lossy (e.g. polaritonic)} \\              
        \frac{\omega_0}{\Delta\omega}\frac{\left[\chi(\omega)\right]^2}{\chi(\omega)+1} & \text{lossless (dielectric)} \\
        \frac{|\sigma(\omega)|^2}{\Re \sigma(\omega)}  & \text{2D materials},
    \end{dcases} \label{eq:fompb} 
\end{equation}
where the 2D material FOM retains its original simple form. For high-index lossless materials:
 \begin{equation}
     f(\omega) \approx \frac{\omega_0}{\Delta\omega} \chi(\omega) = \frac{\omega_0}{\Delta\omega} \left(n^2 - 1\right) \qquad \text{lossless, high-index}. \\
    \label{eq:fompb2} 
\end{equation}		
The material FOM for metals in the small-bandwidth limit is dictated by material loss $\Im{\chi(\omega)}$, whereas the FOM for lossless materials (dielectrics) is dictated by relative bandwidth $\Delta\omega / \omega_0$. Lossy materials simplify to the single-frequency metric $|\chi|^2 / \Im \chi$ because the bandwidth of their resonant response is dictated by material loss, and their single-frequency response can be maintained over the whole bandwidth $\Delta\omega$. The material FOM for lossless high-index materials is consistent with the metrics of \secref{dielectric}, where we saw that all-frequency response is determined by the refractive index; the material FOM of \eqref{fompb2} increases in proportion to $\omega_0$ over $\Delta\omega$, which can be interpreted as the possibility for resonant amplification of the typical $n^2$ enhancement in a resonator with $Q$-factor given by $\omega_0 / \Delta\omega$. No assumption of single-mode or quasistatic behavior is made in \eqref{fompb}, which holds for any number of resonances as well as more complex phenomena such as Fano interactions~\cite{Fano1961,Limonov2017} and exceptional points~\cite{Kato2013,Heiss2004}.

\Figref{bandwidthFOM} compares the material FOM for a large variety of materials at optical frequencies. On the left side of the figure is the material FOM for canonical material types: (a) a Drude metal, $\chi(\omega) = -\omega_p^2 / (\omega^2 + i\gamma\omega)$, for plasma frequency $\omega_p$ and loss rate $\gamma$, (d) a lossless, constant-susceptibility ($\chi(\omega) = 9$) material, and (g) a ``Drude'' 2D material, with conductivity $\sigma(\omega) = i\omega_p / (\omega + i\gamma)$. One can see that these three material types show very different characteristic dependencies of their FOM on frequency and bandwidth. The Drude-metal FOM is nearly independent of bandwidth for small-to-moderate bandwidths and increases with the center wavelength (of the frequency band of interest), $\lambda_0$. On the other hand, a dielectric with constant permittivity is independent of center wavelength, and highly dependent on the bandwidth. Finally, 2D Drude conductors are somewhere in between. Loss originates from both the material parameter $\gamma$ as well as the bandwidth, and the material FOM favors small bandwidth and large wavelength (for a large conductivity). These simplified permittivity/conductivity models describe the key features of the FOM for real materials, as the plots in \figref{bandwidthFOM}(b,c,h) follow the same trends as those in \figref{bandwidthFOM}(a,d,g): the material FOM for metals increases with wavelength, whereas dielectrics (Si and SiC) and polaritonic materials (SiO$_2$ and TiO$_2$) that support surface-phonon polaritons at mid-IR frequencies~\cite{Maier2007} are relatively constant with wavelength. Conversely, the plots in \figref{bandwidthFOM}(e,f,i) show the effects of increasing bandwidth, with the FOM values nearly unchanged for metals but those of the dielectrics and polaritonic materials decreasing nearly linearly. The material FOM of 2D conductors increases with both wavelength and smaller bandwidths.

\section{Looking Forward}
\label{sec:forward}
The results highlighted in this paper demonstrate a synergy between experimental materials discovery and theoretical nanophotonic bounds. There are now a number of key metrics by which materials can be evaluated for optical performance. Looking forward, these results prompt new questions in a variety of directions. First, the metrics can drive new-material synthesis, whether for moderative-negative-permittivity polaritonic material with particularly small loss (as measured by $\Im \chi / |\chi|^2$) or for dielectric materials with particularly large refractive indices $n$ and small chromatic dispersion. From a theoretical perspective, a quantum-mechanical analysis may suggest constituent atoms or alternative approaches to achieving anomalously large material figures of merit, or they may provide insight through novel bounds on how large such metrics can be. There is also the question of which materials are optimal for quantum-photonic \emph{applications}, where high radiative efficiency (as discussed in \secref{radeff}) is important, and a number of considerations beyond maximum response may be desirable. This highlights another area for exploration---alternative nanophotonic metrics. Beyond maximum response, quantities such as nonreciprocal transmission, isolation, selectivity, and others may be desirable~\cite{Shen2014,Tzuang2014,Yu2015,Shi2015}; bounds on such response functions may introduce new material metrics for such scenarios. \emph{Active} nanophotonic platforms offer another area for exploration, with metrics such as switching speed taking increased importance. These examples provide a glimpse at the fertile opportunity for better understanding of the extreme limits of light--matter interactions.

\end{document}